\documentclass[lettersize,journal]{IEEEtran}
\usepackage{amsmath,amsfonts}
\usepackage{algorithmic}
\usepackage{algorithm}
\usepackage{array}
\usepackage[caption=false,font=normalsize,labelfont=sf,textfont=sf]{subfig}
\usepackage{textcomp}
\usepackage{stfloats}
\usepackage{url}
\usepackage{verbatim}
\usepackage{graphicx}
\usepackage{cite}
\usepackage{svg}

\usepackage{array} 
\usepackage{hyperref}
\usepackage{xurl}
\hypersetup{
  colorlinks=true, 
  linkcolor=blue,   
  citecolor=blue,  
  urlcolor=blue     
}
\usepackage{booktabs} 

\usepackage{float} 

\usepackage[belowskip=-1pt,aboveskip=0pt]{caption}
\usepackage{subcaption}

\usepackage[sort&compress,numbers]{natbib} 

\begin{document}

\title{Maximizing Grid Support of Electric Vehicles by Coordinating Residential Charging: Insights from an Arizona Feeder Case Study}

\author{Mohammad Golgol}
\author{Anamitra Pal}
\author{Vijay Vittal}

\author{\IEEEauthorblockN{Mohammad Golgol\IEEEauthorrefmark{1}, \textit{Student Member, IEEE}, Anamitra Pal\IEEEauthorrefmark{1}, \textit{Senior Member, IEEE}, Vijay Vittal\IEEEauthorrefmark{1}, \textit{Life Fellow, IEEE}, Christine Kessinger\IEEEauthorrefmark{2}, Ernest Palomino\IEEEauthorrefmark{2}, \textit{Member, IEEE}, and Kyle Girardi\IEEEauthorrefmark{2} \\
\IEEEauthorblockA{\IEEEauthorrefmark{1}School of Electrical, Computer, and Energy Engineering, Arizona State University, Tempe-85287, AZ, USA} \\
\IEEEauthorblockA{\IEEEauthorrefmark{2}Salt River Project (SRP), 6504 East Thomas Road, Scottsdale-85251, AZ, USA\\
}
}

\thanks{This work was supported in part by the SRP/ASU Joint Research Project No. EE-014.

Emails: \url{mgolgol@asu.edu}; \url{Anamitra.Pal@asu.edu}; \url{vijay.vittal@asu.edu}; \url{Christine.Kessinger@srpnet.com}; \url{Ernest.Palomino@srpnet.com}; \url{Kyle.Girardi@srpnet.com}}
}

\markboth{Under Peer Review}%
{Shell \MakeLowercase{\textit{et al.}}: A Sample Article Using IEEEtran.cls for IEEE Journals}


\maketitle

\begin{abstract}
The installation of high-capacity fast electric vehicle (EV) chargers at the residential level
is posing a significant risk to the distribution grid. This is because
the increased demand 
from such forms of 
charging 
could exceed the ratings of the distribution assets, particularly, transformers.
Addressing this issue is critical, given that current infrastructure upgrades to enhance EV hosting capacity are both \textit{costly} and \textit{time-consuming}. 
This study addresses this challenging problem by introducing
a novel
algorithm to maximize residential EV charging without overloading any transformer within the feeder.
The proposed method is applied to a real-world utility feeder in Arizona, which includes 120 transformers of varying capacities. The results demonstrate that this approach effectively manages a substantial number of EVs without overloading 
the transformers.
It also identifies locations that must be prioritized for future upgrades. The proposed
framework can serve as a valuable reference tool for utilities when conducting distribution system planning 
for supporting the growing EV penetration.
\end{abstract}
\vspace{-2 pt}
\vspace{-8 pt}
\section{Introduction}
\label{sec:introduction}

\IEEEPARstart{I}nternal combustion engine (ICE)
vehicles have traditionally been
one of the major contributors to greenhouse gases and air pollution \cite{mehlig2023accelerating}. In this regard, electric vehicles (EVs) offer a promising solution as they reduce
the carbon
footprint of transportation.
The value proposition is so high that most major car manufacturers have started developing EVs
of their own design \cite{hemavathi2022study}.  
This expected rise in EV production is prompting power utilities to prepare for a significant uptick in EV ownership. 
For instance, Salt River Project (SRP), a power utility in Arizona, \textit{anticipates supporting approximately one EV for every two households by 2035}.

In addition to environmental consciousness, this surge in EV adoption is also being fueled by government incentives and declining battery prices, which has resulted in nearly a doubling of the number of personal EVs in recent years \cite{iea_ev}. At the same time, since 84\% of EV owners in the U.S. charge their vehicles at home \cite{jd_power_ev}, this trend underscores the importance of coordinating
residential charging in the distribution system
and the need for robust hosting capacity (HC) evaluations. 
In the context of this paper, 
HC refers to the maximum EV load that can be accommodated without violating operational constraints \cite{sandstrom2023evaluation}, usually in the form of transformer overloading.

Based on an extensive data-driven analysis, \cite{yu2022data} highlighted the vulnerability of low-voltage grids to transformer overloading under high EV penetration scenarios, and emphasized the need for targeted infrastructure upgrades.
Similarly, \cite{roy2023impact} demonstrated through multi-physics reliability assessments that increased EV penetration accelerates transformer lifetime deterioration, with all the potential adverse impacts intensifying 
during peak demand periods.
To partially meet the EV charging load demands, \cite{rabiee2021enhanced} utilized renewable energy resources as an additional power source. 
However, the intermittent nature of renewable resources, unavailability during peak demand periods (evening hours), and uncertainties related to EV load profiles lower the practicality of this approach 
for widespread EV integration.

To manage the uncertainties and risks associated with EV charging, \cite{aliasghari2020risk, nikkhah2020stochastic, geng2021coordinated, ali2019voltage,mulenga2021adapted,najafi2021optimal} proposed a variety of optimization formulations. 
However, these formulations 
often relied on static assumptions and/or 
did not 
fully 
capture the interactive effects among the various parameters that influence the charging process \cite{mousa2024comprehensive}.
Specifically, parameters such as \textit{state-of-charge (SOC), charger power ratings, commuting patterns, and customer demographics} interact with each other,
and are collectively responsible for shaping the real-world load profiles. All of these factors must be considered in the optimization formulation to ensure realistic outcomes.


\textcolor{black}{To account for customer behavior and demand response, \cite{clement2009impact, leemput2014impact, lopes2024demand} generated EV charging profiles by assuming \textit{uniformly distributed} charging sessions throughout the day.} 
Similarly, \cite{veldman2014distribution, elnozahy2013comprehensive, leou2013stochastic, zhu2022assessing} attempted to model EV charging profiles by incorporating travel surveys of conventional vehicles to account for user behavior. However, they made static assumptions by considering \textit{only the most expected scenario}. Such simplifying assumptions regarding charging sessions, along with the overlooking of individual variations in user preferences and commuting patterns, limit the validity of the outcomes.

For instance, as highlighted in \cite{zhang2020daily}, accurately modeling behavioral patterns, such as departure and return times, is crucial for ensuring the robustness of the analysis. Failing to account for the diversity in EV user preferences can lead to underestimating or overestimating system capabilities and their corresponding impact on distribution transformer health.

Lastly, studies such as \cite{yu2022data} evaluated the impact of incorporating different levels of EV penetration on distribution transformers, but the analysis was restricted to a single type of transformer. 
In order to comprehensively assess HC of a distribution feeder, the impact of EV loads on all the feeder transformers must be analyzed. 

This paper seeks to address the knowledge gaps identified above
by proposing a holistic, end-to-end framework that integrates real-world advanced metering infrastructure (AMI) data gathered from an actual distribution feeder, which includes 120 transformers, along with detailed commuting patterns and realistic EV charging scenarios. 
From a \textit{planning perspective}, this comprehensive approach is essential for confidently assessing system capabilities, identifying necessary upgrades, and prioritizing locations to mitigate potential overloading conditions.
The key contributions of this work are as follows:
\begin{itemize}
    \item A framework is introduced to comprehensively evaluate the HC of distribution feeders, offering a practical solution for system planners of power utilities.

    \item The framework relies on real AMI data, ensuring accurate assessments of system behavior under various scenarios, as opposed to relying on
    assumed distributions.

    \item The coordination model takes into account key factors such as EV availability, charger types, and peak demand. It optimizes charging schedules to minimize costs to customers and considers their availability while preventing overloading of the associated transformer.

    \item The study provides utilities with a method to determine the maximum number of EVs that each transformer can reliably support, along with a confidence rate to account for the stochastic nature of the problem, giving a clearer understanding of system reliability and grid flexibility.

\end{itemize}

By addressing these crucial areas, this work offers utilities a scalable and robust framework for assessing and managing EV hosting capacity in real-world distribution systems.

The rest of the paper is organized as follows: 
Section  \ref{problem_back} discusses the parameters that are critical to modeling an EV load profile.
Section \ref{prop_formula} gives a detailed description of the developed optimization models. Section \ref{ImpProc} presents the proposed end-to-end framework for assessing all transformers within a given distribution system. Section \ref{Results} showcases the effectiveness of the proposed method and discusses insights derived from the results.
The conclusions are provided in Section \ref{Conclusion}.

\vspace{-10 pt}
\section{Problem Background}\label{problem_back}
\subsection{Type of Chargers}
\label{subsec:TypeOfCharger}
The rapid growth in EV adoption has brought about a critical need to understand the various types of available EV chargers, especially at the residential level. Level 1 chargers are the most basic type of EV chargers, and use a standard 120-volt AC outlet \cite{zeb2020optimal}. These chargers typically deliver between 1.4 kW and 1.9 kW of power, making them suitable for overnight charging. Due to their very slow charging speed, such types of chargers do not provide much flexibility in EV charging coordination.

Level 2 chargers operate on a 240-volt AC outlet, similar to what is used for large household appliances. 
These chargers are capable of delivering from 2.5 kW to 19.2 kW of power, significantly reducing the time required to charge an EV compared to Level 1 chargers. Level 2 chargers are the most common choice for residential charging due to their balance of cost, installation requirements, and charging speed \cite{sayed2022review}.
These chargers typically come in slow (3.6 kW) and fast (7.2 kW) varieties. Slow chargers are  more suitable for plug-in hybrid EVs (PHEVs), while fast chargers are more suitable for fully electric cars.

The efforts to further facilitate Level 2 charging at residential complexes has led to the introduction of a home charger called the Wall Connector by Tesla, which offers a range of charging powers from 2.8 kW to 11.5 kW \cite{Wall_connector}. This charger allows users to tailor charging speeds to their specific needs and electrical capabilities. Additionally, the Wall Connector is compatible with other types of EVs. The flexibility provided by devices such as the Wall Connector is critical because the impact of EV chargers on the distribution system is heavily influenced by the charging power.
\vspace{-10 pt}
\subsection{Commuter Departure and Return Patterns}\label{subsec: CommuterPattern}
The travel behaviors of EV users and the precision of the probability distribution models that describe these behaviors are pivotal in forecasting the charging demands placed on the power grid. Traditional optimization-based coordination models for EV charging, as discussed in \cite{aliasghari2020risk, nikkhah2020stochastic, geng2021coordinated, ali2019voltage,mulenga2021adapted,najafi2021optimal,dominguez2023single,nazari2020electric, quiros2015control, xu2015hierarchical}, attempted to capture such individual commuting attributes in their models by either making uniform static assumptions or using the most likely statistical parameters. However, they often fail to adequately consider the real-world connection and disconnection times that reflect the convenience/preferences of EV owners. This gap can result in scenarios where \textit{EV owners are unable to charge their vehicles at the suggested times} or \textit{opt to charge during peak demand hours due to personal energy needs}. Such misalignments introduce uncertainties regarding the effectiveness of coordination schemes in preventing grid overloads as EV adoption increases.

To enhance accuracy of the proposed scheme, real-world travel pattern data is incorporated from the National Household Travel Survey (NHTS) \cite{NHTS}. The NHTS provides detailed travel data for each vehicle, including trip start and end times, driving duration, distance, and parking duration. Additionally, it offers demographic information about vehicle owners, such as age, gender, education, and income, enabling a nuanced analysis of travel behaviors across different user profiles.
Prior research has demonstrated that demographic factors such as age, education, and income significantly influence daily travel and energy consumption patterns \cite{zhang2020daily}. Therefore, use of the NHTS dataset is critical in ensuring realism of the study.

\begin{figure}[t]
\centerline{\includegraphics[width=0.485\textwidth]{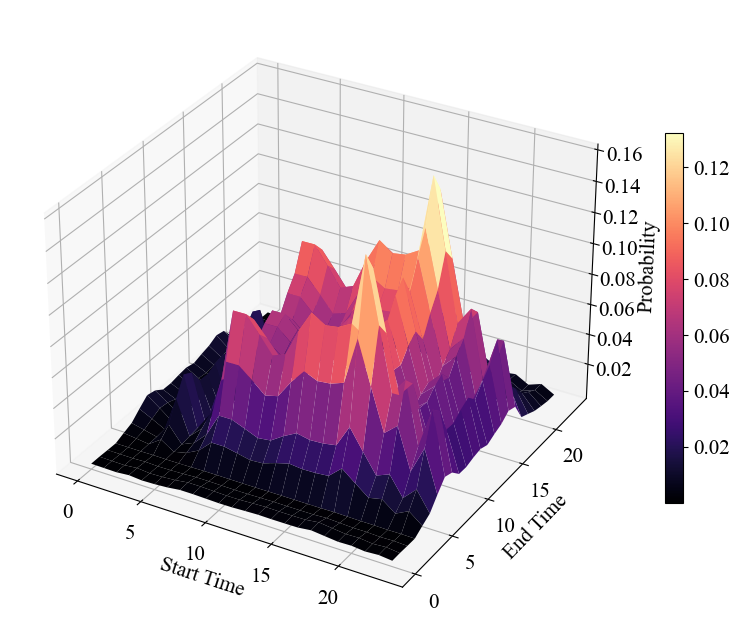}}
\caption{Joint PMF of trip Start and End Times for EV commuters}
\vspace{-10 pt}
\label{fig: DailyCom}
\end{figure}

In this study demographic information is used in the following manner. 
Taking age as a representative demographic factor for EV owners and using data on trip start and end times from the NHTS database, a joint probability mass function (PMF) is developed, as depicted in Fig. \ref{fig: DailyCom}.
The resulting PMF reveals that travel patterns vary significantly and are distributed throughout the day, rather than being confined to specific time intervals. \textcolor{black}{Incorporating these insights into the optimization formulation enables the development of more effective and reliable EV charging coordination models that align more closely with the actual usage patterns of EV owners.}

\subsection{Type of EV}\label{subsec: TypeOfEV}
Given the different types of EVs, their battery capacity plays a significant role in assessing the impacts on the power grid. Fully electric cars typically have larger battery capacities to ensure a longer electric driving range, ranging from 30 kWh to over 100 kWh. These vehicles are designed to operate solely on electricity,
relying entirely on their battery packs for propulsion. For instance, the Tesla Model 3 Long Range has a battery capacity of around 75 kWh \cite{TeslaModelY}, the Tesla Model S Long Range can have up to 100 kWh \cite{TeslaModelS}, while the Nissan Leaf offers battery capacities ranging from around 40 kWh to 62 kWh \cite{NissanLeafSpec}.
In contrast, PHEVs generally have smaller battery capacities, ranging from 8 kWh to 20 kWh. 
For example, the Toyota Prius Prime has a battery capacity of about 13.6 kWh, while the RAV4 Prime has a battery capacity of around 18.1 kWh \cite{ToyotaPHEVs}.

The proportion of each type of EV significantly impacts the assessment of a transformer's load, as the required power to recharge an EV can vary greatly. At the same time, adoption rates and market shares highlight the competitive landscape and consumer preferences. As of October 2023, Tesla dominated the U.S. market with a 56\% share, followed by Chevrolet and Ford with 5.9\% and 5.8\%, respectively \cite{TypeOfEV}. However, determining the exact types of EVs connected to a specific transformer can be challenging when assessing an entire feeder.
Following the power system principle of planning for the \textit{most realistic worst-case scenario}, it is assumed for the case study that all EVs are Tesla Model S vehicles with a battery capacity of 100 kWh. Considering this scenario ensures that the conclusions drawn from the analysis will hold under practical circumstances.

\vspace{-8 pt}
\subsection{State of Charge (SOC)}\label{subsec: SOC}
The SOC is another critical component in modeling the charging behavior of EVs. It represents the current level of charge in the vehicle's battery as a percentage of its total capacity. It directly affects the EV users' charging requirements as lower SOC levels often necessitate longer charging sessions and higher power demands, which can significantly influence the total load on the power distribution system. At the same time, charging up to 100\% is typically reserved only for occasions when maximum range is needed for longer trips, and not for daily commute.

Since all EVs in this study are assumed to be Tesla Model S, according to Tesla's maintenance guidelines, it is advisable to recharge the battery before it falls below 20\% \cite{TeslaBattery}. These practices are based on the operational characteristics of Lithium-ion batteries, where extreme charge levels can stress the cells.
Therefore, in this study which focuses on regular use, the goal is set to maintain battery SOC between 20\% and 80\%. Two different distributions are employed in this study to model SOC:
\begin{itemize}
    \item \textit{Initial SOC}: A uniform distribution is assumed, ranging between 20\% and 30\%, representing the typical SOC when EV owners begin charging.
    The choice of a uniform distribution stems from the fact that any SOC value between this range has the same likelihood of being selected as the starting value for charging.
    Lastly, note that with the Tesla Model S’s energy consumption, including charging losses, estimated at 122 MPGe (or about 276 watt-hours per mile, equivalent to 3.6 miles per kWh), this range corresponds to the energy required for a commute of 72 to 108 miles, which is still sufficient for most daily commutes.
    \item \textit{Final SOC}: A Chi-Squared distribution is used, peaking near 80\% and tapering off towards 100\%, reflecting the SOC levels typically preferred by EV owners after charging is completed.
    The choice of a Chi-Squared distribution stems from the fact that EV batteries have optimal performance when regularly charged till 80\%, with charging rates slowing down beyond this value. Similarly, a full charge is usually preferred when a longer trip is planned, which does not happen on a daily basis.
\end{itemize}

By using these distributions, the charging power for each EV is modeled realistically while also being in alignment with the industry best practices for battery maintenance.

\vspace{-10 pt}
\subsection{Pricing Plan}\label{subsec: Pricing Plan}
Power utilities have implemented various pricing strategies to manage increased load, especially during peak hours. 
One of the most popular strategies is the time-of-use (TOU) plan in which the price of electricity varies depending on the time-of-the-day.
For example, SRP's TOU plan (shown in Table \ref{SRP_TOU_Plan} below) \cite{SRPPricingPlan} divides the day into peak and off-peak periods and encourages its customers to operate heavy-load appliances during off-peak hours.
In essence, the TOU plan accommodates higher residential demands while promoting energy conservation and cost efficiency.

\begin{table}[ht]
\centering
\caption{SRP TOU Price Plan}
\label{SRP_TOU_Plan}
\resizebox{0.485\textwidth}{!}{\begin{tabular}{cccc}
\toprule
\textbf{Season} & \textbf{Months} & \textbf{Time Period} & \textbf{Prices per kWh} \\ 
\hline
Winter & Nov. through Apr. & 5 AM - 9 AM  & 11.45¢ \\ 
\cline{3-4}
 &  & 9 AM - 5 PM & 8.85¢\\ 
\cline{3-4}
 &  & 5 PM - 9 PM  & 11.45¢\\ 
 \cline{3-4}
 &  & 9 PM - 5 AM & 8.85¢\\ 
 \cline{3-4}
\hline
Summer & May / June / Sept. / Oct. & 2 PM - 8 PM  & 22.70¢ \\ 
\cline{3-4}
 &  & 8 PM - 2 PM  & 9.03¢\\ 
\cline{3-4}
\hline
Summer Peak & July / Aug. & 2 PM - 8 PM  & 25.85¢\\ 
\cline{3-4}
 &  & 8 PM - 2 PM & 9.06¢\\ 
\bottomrule
\end{tabular}
}
\end{table}

It can be observed from the table that
SRP's pricing plan varies significantly between the winter and peak-summer months which reflects the seasonal energy demand dynamics of Arizona. 
From November to April, rates are structured into three segments: a peak rate of 11.45 \textcent $ $ per kWh from 5 AM to 9 AM and 5 PM to 9 PM, while 8.85 \textcent $ $ per kWh is charged for the other hours.
Conversely, during the peak summer months (July and August), when energy consumption typically spike due to very high air conditioning usage, a more aggressive rate structure is implemented. This includes rates of 9.06 \textcent $ $ per kWh from before 2 PM to after 8 PM, and a very high peak rate of 25.85 \textcent $ $ per kWh from 2 PM to 8 PM.

\vspace{-10 pt}
\section{Proposed Optimization Formulation}\label{prop_formula}
In this section, an optimization model is designed to establish an effective charging schedule for EVs that minimizes costs while adhering to specific safety limitations and requirements. 
The primary concern is the potential strain on transformer capacity in residential areas, which is highly likely to be exceeded when multiple EVs are connected simultaneously for charging. Additionally, the cost of charging is a critical factor from the consumer’s perspective. Therefore, the proposed model aims to coordinate the charging process to minimize charging expenses of EV owners based on their availability and
utility pricing plans while ensuring that none of the transformers get overloaded.
The constrained optimization problem is outlined below:

\begin{equation}
    \begin{aligned}
        \min_{\kappa[n,t]} \sum_{n=0}^{|\chi|-1} \left( \sum_{t=0}^{|\tau|-1} \nu \cdot \kappa[n,t] \cdot p[t] \right) 
    \end{aligned}\label{obj}
\end{equation}
subject to:
\begin{equation}
    \begin{aligned}
    \left( \sum_{n=0}^{|\chi|-1} \nu \cdot \kappa[n,t] \right) + \iota[t] \le C_{\max} && \forall t \in \tau
    \end{aligned}\label{cons1}
\end{equation}
\begin{equation}
    \begin{aligned}
    \kappa[n,t-1]-\alpha(1-\omega[n,t])+\beta \le \kappa[n,t] \\ \forall n \in \chi, \text{ and }\forall t \in \{1,...,\tau_{\max}\}
    \end{aligned}\label{cons2}
\end{equation}
\begin{equation}
    \begin{aligned}
        \kappa[n,t - 1] + \alpha \cdot \omega[n,t] \ge \kappa[n,t]  \\ \forall n \in \chi, \text{ and } \forall t \in \{1,...,\tau_{\max}\} 
    \end{aligned}\label{cons3}
\end{equation}
\begin{equation}
    \begin{aligned}
        \omega[n,t] = \kappa[n,t]  &&\forall n \in \chi,  \text{ and } t=0
    \end{aligned}\label{cons4}
\end{equation}
\begin{equation}
    \begin{aligned}
        \kappa[n,0] \le \kappa[n,t] && \forall n \in \chi,  \text{ and } \forall t \in \{0,...,\tau_{\min}\} 
    \end{aligned}\label{cons5}
\end{equation}
\begin{equation}
    \begin{aligned}
         \kappa[n,t] - \kappa[n,t - 1] \le \kappa[n,t + i] \\ \forall n \in \chi, \forall i \in [0, \tau_{\min}], \text{ and } \forall t \in [1, \tau_{\max} - \tau_{\min}]
    \end{aligned}\label{cons6}
\end{equation}
\begin{equation}
    \begin{aligned}
        \kappa[n,t] \ge \kappa[n,t + 1] \\ \forall n \in \chi, \text{ and }\forall t \in \{\tau : \tau \geq (\tau_{\max} - \tau_{\min})\}
    \end{aligned}\label{cons7}
\end{equation}
\begin{equation}
    \begin{aligned}
        \sum_{t=0}^{|\tau|-1} \omega[n,t] &\leq S & \forall n \in \chi
    \end{aligned}\label{cons8}
\end{equation}
\begin{equation}
    \begin{aligned}
        \kappa[n,t] = 0 && \forall n \in \chi, \text{ and }\forall t \in \Delta_{n}
    \end{aligned}\label{cons9}
\end{equation}
\begin{equation}
    \begin{aligned}
        \sum_{t=0}^{|\tau|-1} \left( \nu \cdot \kappa[n,t] \right) \textcolor{black}{=} D_n && \forall n \in \chi
    \end{aligned}\label{cons10}
\end{equation}

The objective function (\ref{obj}) minimizes the total charging cost for all EVs connected to a specified transformer.
\textcolor{black}{The parameter $\nu$ represents the given charging power, and $\kappa[n,t]$ is a binary variable indicating whether the EV $n$ is connected to the grid at time $t$. The price of energy at time $t$ is denoted by $p[t]$ (obtained from Table \ref{SRP_TOU_Plan} in
Section \ref{subsec: Pricing Plan}).}


The constraints that must be met are outlined in (\ref{cons1})-(\ref{cons10}). Here, \(\chi\) represents a set of EVs, and \( \tau\) is a set of time intervals, each with a duration of 15 minutes which results in a total of 96 intervals per day. The parameter $\tau_{\max}$ is fixed at 96, indicating the index of the last interval, while $\tau_{\min}$ specifies the minimum number of consecutive time slots an EV must remain connected, set at 4, which is equivalent to one hour.
Constraint $(\ref{cons1})$ ensures that the combined load of household appliances and EV chargers does not exceed the transformer’s maximum capacity $C_{\max}$, with $\iota[t]$ representing the aggregated household load at time $t$.

Equations (\ref{cons2})-(\ref{cons4}) aim to extend the battery life of EVs by regulating the frequency of changes in the charging status. In these equations, the binary variable \(\omega\) tracks the number of switchings that occur between the two modes (namely, connected or disconnected) of the charging schedules.
This regulation of the charging frequency plays a key role in managing the power drawn from the transformer, which affects the available capacity to support
additional EV charger loads. 
As a result, the available capacity often necessitates connecting or disconnecting EVs from the grid to prevent overloading the transformer. Lastly, the parameters \( \alpha \) and \( \beta \) are two arbitrary positive constants that satisfy (\ref{cons2}) and (\ref{cons3}). Simply put, if \(\kappa[n,t-1]<\kappa[n,t]\), then the binary decision variable \(\omega[n,t]\) must be set to 1. If not, \(\omega[n,t]=0\) should be applied to uphold these constraints.

Equations $(\ref{cons5})$ to $(\ref{cons7})$ ensure that there is no change in the charging status if the connection duration is less than one hour. Specifically, $(\ref{cons6})$ requires that each connection lasts for more than four time slots, while $(\ref{cons7})$ manages specific boundary conditions to prevent switching or initiating a new charging session if the remaining time slots are fewer than $\tau_{\min}$ intervals.
Constraint $(\ref{cons8})$ restricts the total number of switches so that it does not exceed a specified limit, $S$. 
This parameter controls the trade-off between the flexibility of the coordination model and battery longevity. A larger value leads to a more sparse charging schedule, while a smaller value results in longer charging sessions which is better for battery health. However, if this value is too small, it increases the chances of infeasibility.

Equation $(\ref{cons9})$ 
incorporates the daily commute patterns of EV owners into the proposed formulation. In (\ref{cons9}), $\Delta_n$ represents the time slots when an EV, specified by subscript $n$, is not available at home and thus, cannot be connected for charging. By including this equation in the constraints, one ensures that there are no scheduling conflicts with the owner’s commuting times, leading to a seamless charging schedule.
The final constraint, $(\ref{cons10})$, is designed to ensure that all EVs meet their required final SOC by the end of the charging session. Here, $D_n$ represents the demand for each EV, identified by subscript $n$.

The proposed formulation effectively integrates various practical considerations, such as transformer capacity limits, cost, and the daily routines of EV owners, to optimize the residential charging schedule.

\section{Implementation Procedure}
\label{ImpProc}

\subsection{Feeder Evaluation}
\begin{figure*}[ht]
\centerline{\includegraphics[width=0.9\textwidth]{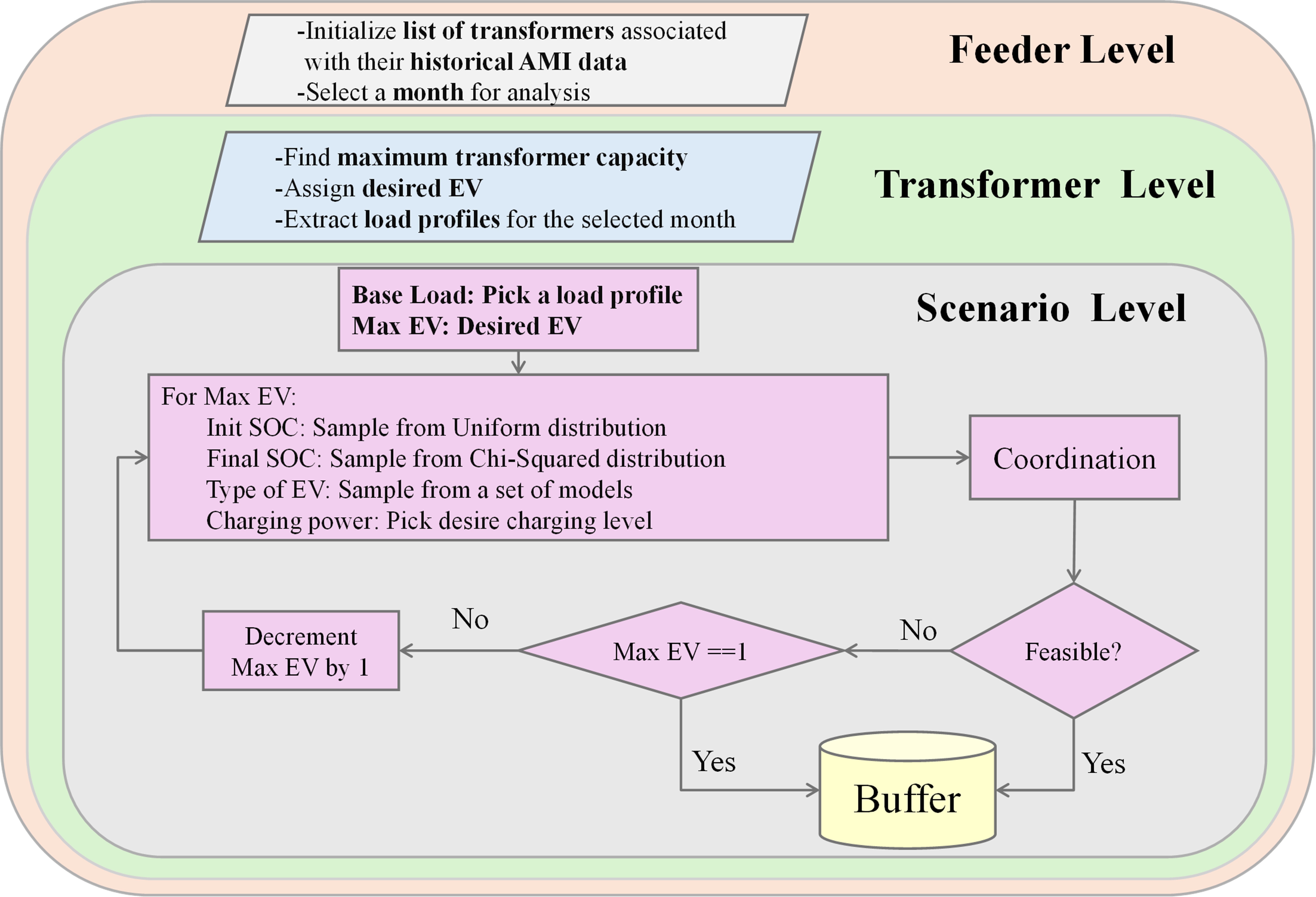}}
\vspace{2mm}
\caption{\textcolor{black}{Proposed methodology for evaluating EV hosting capacity in distribution feeders}}
\label{fig: methodology}
\end{figure*}

This subsection outlines the methodology (depicted pictorially in Fig. \ref{fig: methodology}) that is used to evaluate a distribution feeder's EV hosting capacity.
The procedure is segmented into three distinct levels: \textit{Feeder}, \textit{Transformer}, and \textit{Scenario}, with each representing a progressive stage in the evaluation process. The evaluation begins at the Feeder Level, where each transformer within the feeder is initialized with its corresponding historical consumption data derived from household AMI records.
. A specific month is selected for the evaluation to account for the variability in consumption levels across different days. This stage sets the foundation for more detailed analyses in the subsequent levels.

At the Transformer Level, the focus narrows down to the individual transformer. For each transformer, essential attributes such as its rated capacity and the maximum number of EVs it should support are determined. Additionally, daily load profiles for the designated month are extracted. This process is repeated for every transformer in the feeder.

The Scenario Level involves detailed exploration of specific load profiles, which begins by assigning a base load and estimating the maximum feasible number of EVs (Max \#EV). For each potential scenario, the initial and final SOC are sampled (as detailed in Section \ref{subsec: SOC}), and a type of EV is selected from a predefined set (Tesla Model S for this case study). While the algorithm can handle various charging powers, this case study focuses on two specific (charging power) levels: 7.2 kW and 11.5 kW. Subsequently, the coordination mechanism outlined in Section \ref{prop_formula} is implemented to obtain the optimal EV charging schedule.
 
If a feasible solution is found, the number of EVs that can be accommodated is recorded. If not, the Max \#EV is lowered
by one, and the process is repeated to explore the feasibility of integrating a reduced number of EVs. This iterative process continues until it is determined whether even a single EV can be added without violating the system constraints.
This structured approach allows for a thorough analysis and determination of the maximum HC for EVs at each transformer, providing valuable insights into the feeder's overall capability to support EV charging demands.

\vspace{-10 pt}
\subsection{Coordination in Practice}

This subsection delves into the practical application of the proposed coordination mechanism which serves as a crucial component not only in feeder evaluation but also in the proactive management of transformer loads by utilities to prevent overloads. The first step involves obtaining specific information from EV owners. These include time for charging, current initial SOC, desired final SOC, and EV battery capacity. If such data are unavailable, appropriate values for them can be obtained from historical data, as explained in Section \ref{problem_back}. The second step involves integrating this information with the day-ahead load forecasts to create likely loading scenarios.
In the third and final step, the proposed optimization formulation is solved to produce the charging schedule. This schedule determines if and when an EV should be charged such that no transformer gets overloaded.

It is evident that deploying this coordination requires some infrastructural changes (e.g., sending timely emails/text messages to customers). However, these modifications are generally more cost-effective than the alternative of upgrading multiple transformers in a feeder.
The proposed strategic approach not only optimizes existing infrastructure but also enhances overall efficiency and reliability of power distribution in the face of growing EV usage.

\section{Simulation Results}
\label{Results}

\subsection{Case Study and Data Preparation} 

This case study evaluates the HC of a distribution feeder managed by SRP, which consists of 120 transformers. As outlined in Section \ref{problem_back}, several parameters significantly influence HC. To model the worst-case scenario, specific values for these parameters are identified. Modern chargers, such as the Tesla Wall Connectors, which provide higher charging powers of 7.2 kW and 11.5 kW, are considered in this analysis.
\textcolor{black}{The value of $S$, indicating number of switches, was set to 4. 
For the case study under consideration, this value of $S$ maintained a balance between \textit{feasibility} and \textit{cost} by limiting switching to a small, battery-friendly number of events per day.}
For modeling infeasible time intervals, the joint PMF developed in Section \ref{subsec: CommuterPattern} was utilized. All EVs in this analysis were assumed to be Tesla Model S vehicles with a battery capacity of 100 kWh, as described in Section \ref{subsec: TypeOfEV}. The initial and final SOC were modeled using uniform and chi-squared distributions, respectively, as discussed in Section \ref{subsec: SOC}. Lastly, this study employed SRP's TOU pricing plan, with further details regarding this plan provided in Section \ref{subsec: Pricing Plan}.

\vspace{-10 pt}
\subsection{Findings from AMI Data Analysis} \label{MarchJuly} 
To thoroughly assess the impact of integrating EVs into the feeder, each month was analyzed individually. The results indicated that \textit{July} had the highest electricity consumption, while \textit{March} recorded the lowest. The latter was due to considerable duration of reverse power flows occurring in the feeder-under-analysis in that month because of the dominating presence of roof-top solar photovoltaic resources. 
Since the load profiles for other months fell between these two extremes, the focus of the study was placed on exploring the load profiles from July and March.

\begin{figure}[ht]
\centering
\includegraphics[width=0.475\textwidth]{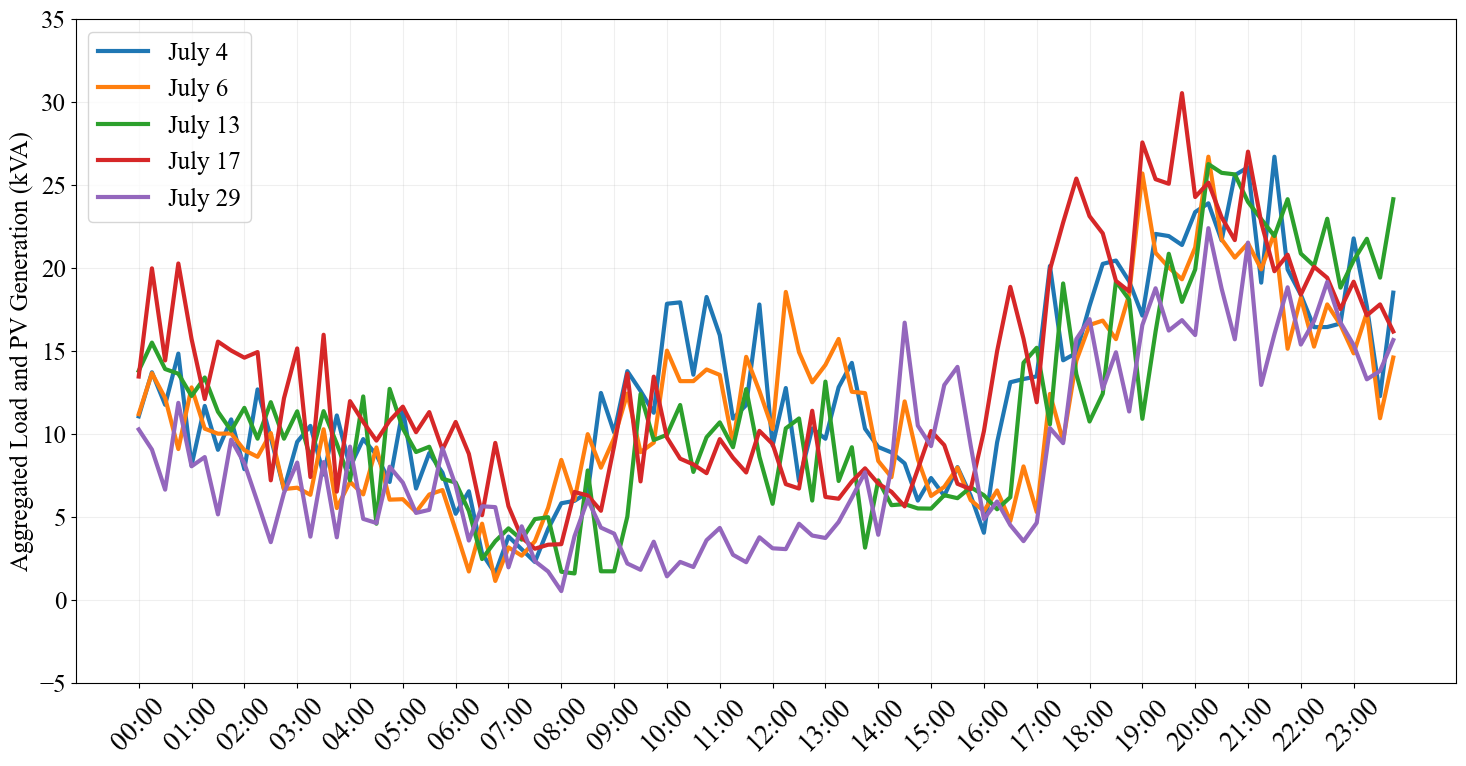}
\caption{Daily profiles of aggregated power drawn from a specific transformer in July with a maximum capacity of 50 kVA}
\label{fig:DailyJuly}
\end{figure}

\begin{figure}[ht]
\centering
\includegraphics[width=0.475\textwidth]{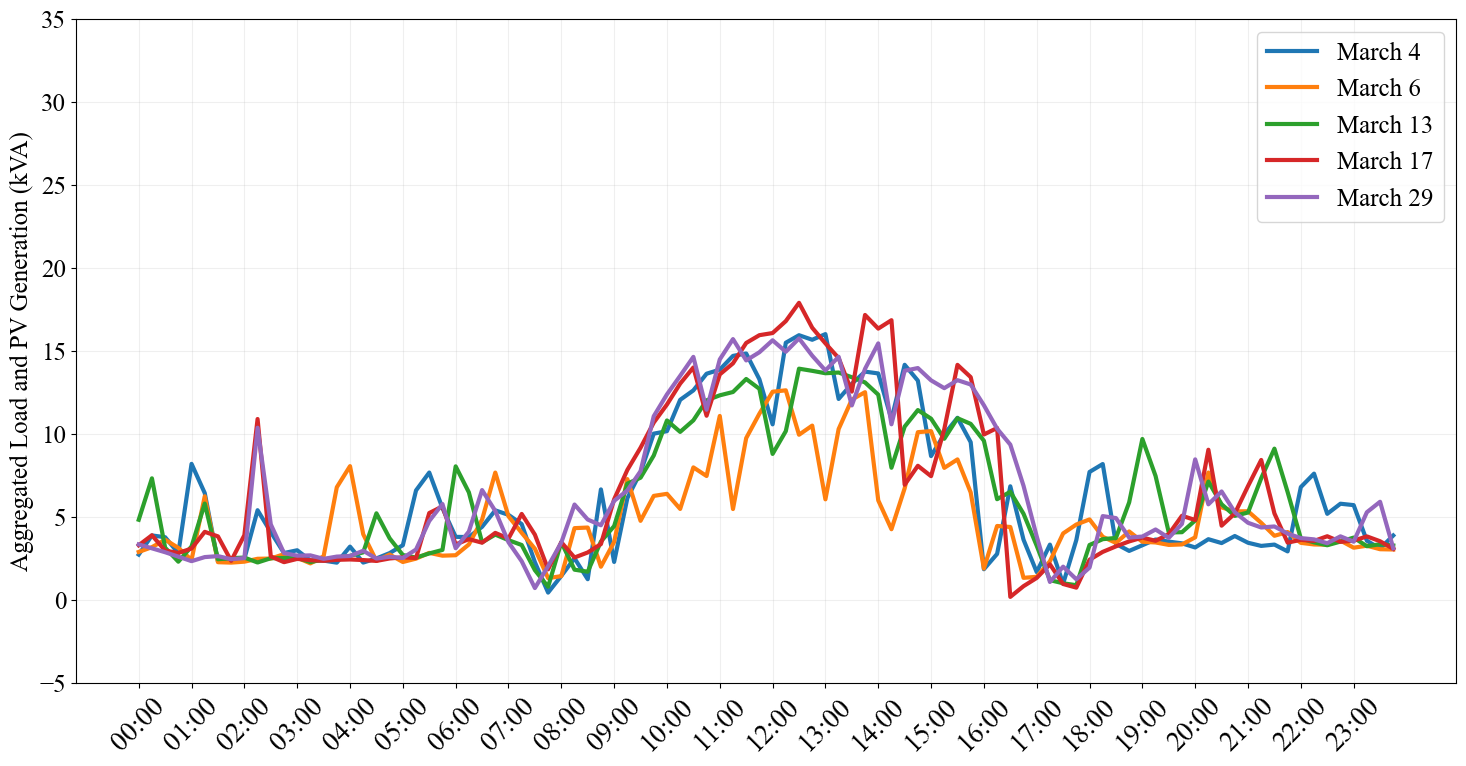}
\caption{Daily profiles of aggregated power drawn from a specific transformer in March with a maximum capacity of 50 kVA}
\vspace{-1em}
\label{fig:DailyMarch}
\end{figure}

Figs. \ref{fig:DailyJuly} and \ref{fig:DailyMarch} display a few load profiles recorded in July and March, respectively. Each curve in these figures represents the total daily power drawn from a selected transformer highlighting peak usage periods. These visualizations also reveal significant fluctuations throughout the day which introduce uncertainty in evaluating the effectiveness of the coordination strategies.
The recorded data shows a general trend of increasing load during the course of the day, with a peak in the afternoon and/or early evening—typical for residential and mixed-use areas due to higher activity during those hours. 
These consumption patterns further justify why minimizing charging costs is the primary objective of the proposed coordination formulation (see Section \ref{prop_formula}). By doing so, not only additional strain on the grid during peak hours is avoided, which is beneficial for utilities, but also it helps prevent higher costs, as owners are likely to shift EV demands to off-peak periods when electricity rates are lower.

\subsection{Performance of Proposed Coordination Scheme under Different Values of Charging Power}
\label{perform_coordination}

\begin{figure}[ht]
\centerline{\includegraphics[width=0.475\textwidth]{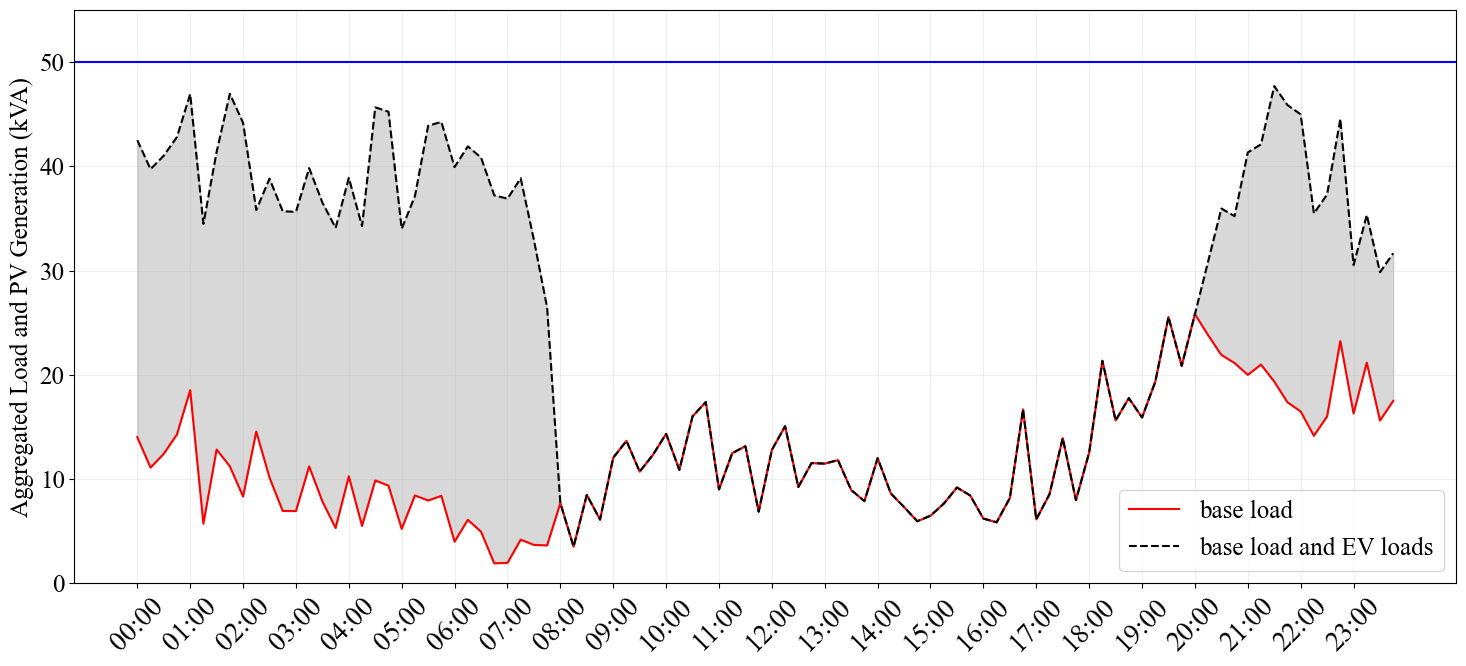}}
\caption{Impact of setting all EV chargers to a fixed power of 7.2 kW on a transformer in July}
\label{fig:out_7.2kW}
\end{figure}

\begin{figure}[ht]
\centerline{\includegraphics[width=0.475\textwidth]{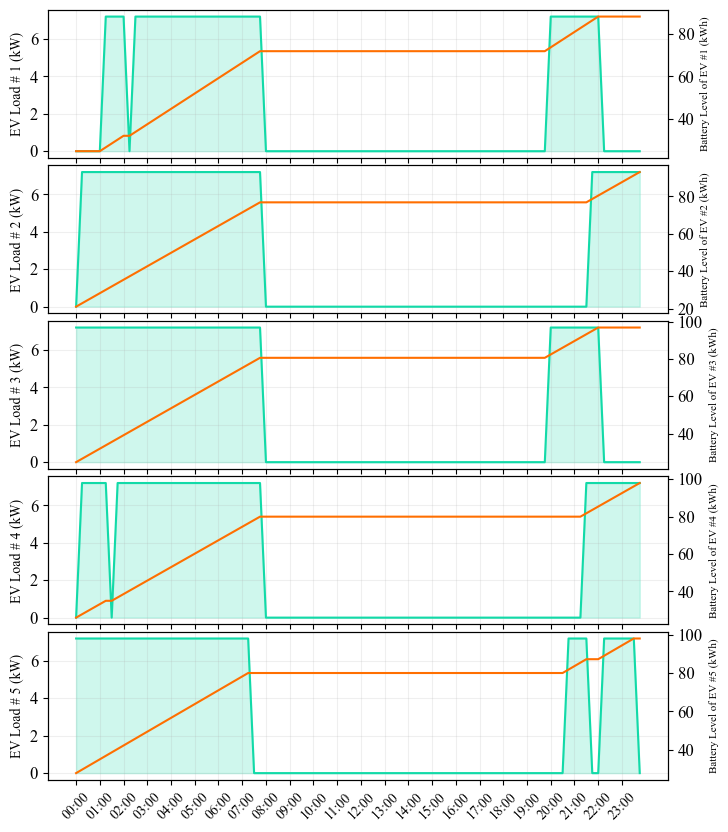}}
\vspace{2mm}
\caption{\textcolor{black}{Coordinated EV charging sessions and battery status at 7.2 kW power. The green color indicates charging duration, while the orange line indicates corresponding battery SoC}}
\label{fig: Char_schedul_7_2}
\end{figure}

\begin{figure}[ht]
\centerline{\includegraphics[width=0.475\textwidth]{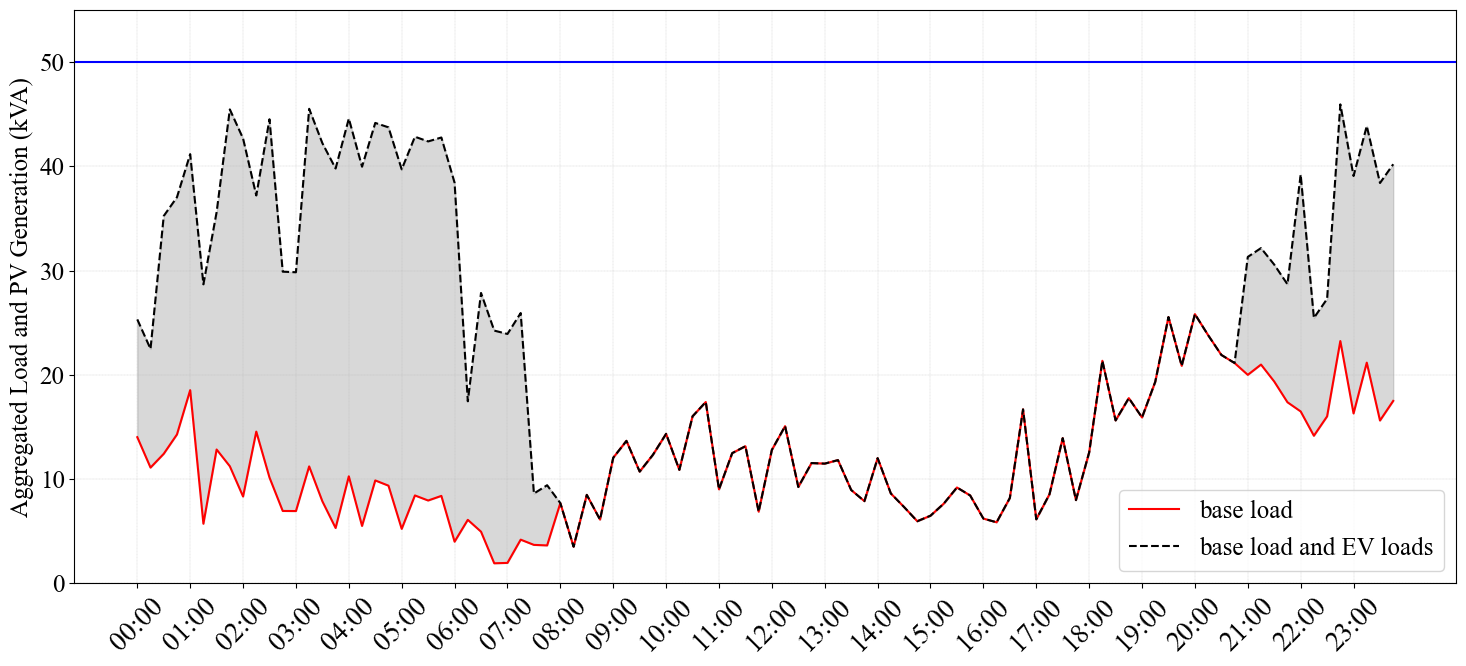}}
\caption{Impact of setting all EV chargers to a fixed power of 11.5 kW on a transformer in July}
\label{fig:out_11.5kW}
\end{figure}

\begin{figure}[ht]
\centerline{\includegraphics[width=0.475\textwidth]{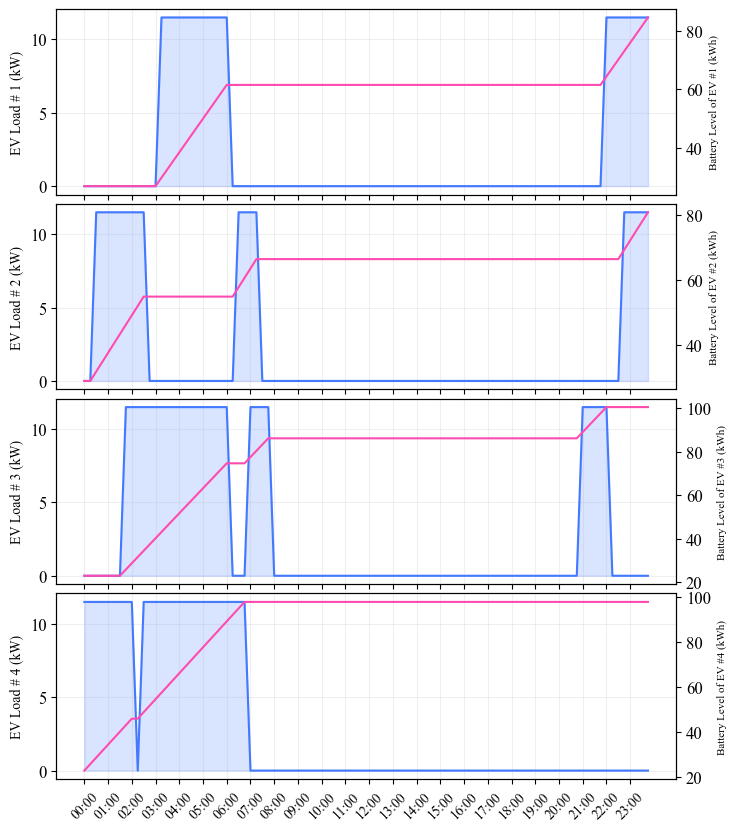}}
\vspace{2mm}
\caption{\textcolor{black}{Coordinated EV charging sessions and battery status at 11.5 kW power. The blue color indicates charging duration, while the pink line indicates corresponding battery SoC}}
\label{fig: Char_schedul_11_5}
\end{figure}

This section showcases the functionality of the proposed coordination method for a given scenario under different levels of charging power; the results are depicted in Figs. \ref{fig:out_7.2kW}, \ref{fig: Char_schedul_7_2}, \ref{fig:out_11.5kW}, and \ref{fig: Char_schedul_11_5}, respectively.
The results correspond to a load profile from July for a transformer with a 50 kVA capacity. The primary objective was to determine the maximum number of EVs that could be supported under two different charging levels: 7.2 kW and 11.5 kW, while all other parameters were fixed. Particularly, it was assumed that the EVs could not be charged between 8 AM and 4 PM (this assumption is relaxed in Section \ref{Compar} for comparison purposes). 

The simulation results, as illustrated in Fig. \ref{fig:out_7.2kW}, analyze the power demand profile by combining the transformer load with the generated charging schedules for five EVs obtained using the proposed coordination scheme.
The associated charging sessions are displayed in Fig. \ref{fig: Char_schedul_7_2}, where five distinct step-shaped curves, labeled EV1 through EV5, illustrate the charging activities at 7.2 kW. Each curve denotes periods of active charging, marked by spikes, and disconnection periods, represented by flat lines at zero level, notably during assumed infeasible time intervals when no charging occurs.
The dashed black line in Fig. \ref{fig:out_7.2kW} refers to the aggregate total demand from both the base load and the EV charger loads. Additionally, the dark green curves in each segment of Fig. \ref{fig: Char_schedul_7_2} indicate the battery levels for each corresponding EV, providing a visual representation of battery consumption and recharge cycles.
\textcolor{black}{It is evident from Figs. \ref{fig:out_7.2kW} and \ref{fig: Char_schedul_7_2} that the coordinated charging strategy not only effectively keeps the overall demand below the transformer’s maximum capacity (denoted by the solid blue line at 50 kVA in Fig. \ref{fig:out_7.2kW}), but also delivers the required energy to bring the battery levels to the assigned final SOC (of at least 80\% as seen in Fig. \ref{fig: Char_schedul_7_2}).}

For the 11.5 kW charging power scenario, similar patterns and outcomes were observed, and they are depicted in Figs. \ref{fig:out_11.5kW} and \ref{fig: Char_schedul_11_5}.
In Fig. \ref{fig: Char_schedul_11_5}, the charging sessions for each of the four EVs are characterized by more compressed charging windows due to the increased (charging) power.

The comparison between the two charging scenarios -- 7.2 kW and 11.5 kW -- reveals a clear relationship between the charging power and the number of EVs that can be accommodated. At the lower power level of 7.2 kW, it is possible to support a higher number of EVs compared to the higher power level of 11.5 kW (five vs. four).
It is also observed that although a lower charging power allows for more EVs to be connected simultaneously, it also necessitates longer charging duration. Furthermore, the analysis indicates that the assumed unavailability of vehicles from 8 AM to 4 PM influences the results. If the non-availability period were shorter, the outcomes could differ markedly. 
This aspect is explored in Section \ref{Compar} where actual commuting patterns are incorporated into the coordination model.

\subsection{Feeder-level Evaluation}
\label{Feeder}

In this sub-section, a comprehensive overview of the impact of integrating a substantial number of EV chargers into the distribution system is provided.
A total of 120 transformers with different capacity within the feeder were evaluated. To effectively summarize the extensive simulation data, the transformers were classified into two main capacity groups: 50 kVA and 75 kVA. However, as transformers within each group supply varying numbers of customers, another challenge arose
in selecting among transformers with the same capacity and customer base. To address this issue and to capture the range of possible outcomes, transformers that exhibited the best and worst results within each capacity group were selected for detailed analysis.

For each selected transformer, 1,500 distinct scenarios were simulated for each month and charging power which amounts to a total of 3,000 simulations per transformer. These scenarios encompassed a variety of household energy consumption patterns, EV charging demands, and other influencing factors as detailed in Section \ref{problem_back}. This approach not only considers the impact of each parameter but also models the interactive effects among them. Additionally, the \textit{desired number} of EVs for each transformer was set to half the number of households it serves which aligns with SRP's projection of one EV for every two houses by 2035.

The simulation results are summarized in Tables \ref{tab: July_Cap50} and \ref{tab: July_Cap75} for July, and Tables \ref{tab: March_Cap50} and \ref{tab: March_Cap75} for March, respectively. Each table presents the outcomes categorized into three columns for each charging power:
\begin{itemize}
    \item \textbf{Infeasibility:} percentage of scenarios where no valid coordination strategy was found to charge even one EV.
    \item \textbf{Less \#EV:} percentage of scenarios where the transformer could supply some EV charging demands but not the desired number of EVs.
    \item \textbf{Desired \#EV:} percentage of scenarios where the transformer successfully hosted the desired number of EVs.
\end{itemize}

\begin{table*}[ht]
\centering
\caption{Comparative analysis of EV charging impact at 7.2 kW and 11.5 kW of power across transformers with 50 kVA capacity in July}
\label{tab: July_Cap50}
\renewcommand{\arraystretch}{1} 
\resizebox{0.8\textwidth}{!}{
\begin{tabular}{lccccccccc}
\toprule
Trans Code & \# Cust & \# EV & \multicolumn{3}{c}{Charging 7.2 kW (\%)} & \multicolumn{3}{c}{Charging 11.5 kW (\%)} \\
\cmidrule(lr){4-6} \cmidrule(lr){7-9}
& & & Infeasibility (\%) & Less \#EV & Desired \#EV & Infeasibility & Less \#EV & Desired \#EV \\
\midrule
$T_{6}$ & 6 & 3 & 0 & 6.27 & \textbf{93.73} & 0 & 1.33 & \textbf{98.67} \\
$T_{12}$ & 7 & 4 & 0 & \textbf{100} & 0 & 0 & \textbf{100} & 0 \\
$T_{13}$ & 7 & 4 & 0 & \textbf{100} & 0 & 0 & \textbf{100} &0 \\
$T_{14}$ & 8 & 4 & 0 & 10 & \textbf{90} & 0 & 1.4 & \textbf{98.6} \\
$T_{19}$ & 5 & 3 & 0 & 6.93 & \textbf{93.07} & 0 & 0.8 & \textbf{99.2 }\\
$T_{30}$ & 6 & 3 & 0.13 & 6.87 & \textbf{93.13} & 0 & 0.5 & \textbf{99.4} \\
$T_{32}$ & 9 & 5 & 0.13 & \textbf{99.87} & 0 & 0 & \textbf{100} & 0 \\
$T_{37}$ & 10 & 5 & \textbf{53.33} & 46.67 & 0 & \textbf{53.33} & 46.67 & 0 \\
$T_{40}$ & 8 & 4 & 0 & 10.4 & \textbf{89.6} & 0 & 1.93 & \textbf{98.07} \\
$T_{43}$ & 5 & 3 & 0.07 & 7.6 & \textbf{92.33 }& 0 & 2 & \textbf{98} \\
\bottomrule
\end{tabular}
}
\end{table*}

\begin{table*}[ht]
\centering
\caption{Comparative analysis of EV charging impact at 7.2 kW and 11.5 kW of power across transformers with 75 kVA capacity in July}
\label{tab: July_Cap75}
\renewcommand{\arraystretch}{1} 
\resizebox{0.8\textwidth}{!}{
\begin{tabular}{lccccccccc}
\toprule
Trans Code & \# Cust & \# EV & \multicolumn{3}{c}{Charging 7.2 kW (\%)} & \multicolumn{3}{c}{Charging 11.5 kW (\%)} \\
\cmidrule(lr){4-6} \cmidrule(lr){7-9}
& & & Infeasibility (\%) & Less \#EV & Desired \#EV & Infeasibility & Less \#EV & Desired \#EV \\
\midrule
$T_{54}$ & 9 & 5 & 0 & 11.33 & \textbf{88.67} & 0 & 1.27 & \textbf{98.73} \\
$T_{57}$ & 8 & 4 & 0 & 10.8 & \textbf{89.2} & 0 & 1.53 & \textbf{98.47} \\
$T_{58}$ & 11 & 6 & 0 & 13.6 & \textbf{86.4} & 0 & 3 & \textbf{97} \\
$T_{60}$ & 13 & 7 & 0.07 & \textbf{99.93} & 0 & 0 & \textbf{100} & 0 \\
$T_{62}$ & 15 & 8 & 0 & 23.33 & \textbf{76.67} & 0 & 23.33 & \textbf{76.67} \\
$T_{66}$ & 9 & 5 & 0 & 11.93 & \textbf{88.07} & 0 & 1.8 & \textbf{98.2} \\
$T_{69}$ & 10 & 5 & 0 & 13.67 & \textbf{86.33} & 0 & 1.67 & \textbf{98.33} \\
$T_{77}$ & 10 & 5 & 0 & \textbf{100} & 0 & 0 & \textbf{100} & 0 \\
$T_{78}$ & 14 & 7 & 0 & 15.6 & \textbf{84.4} & 0 & 2 & \textbf{97.93} \\
$T_{83}$ & 12 & 6 & 0 & 15.87 & \textbf{84.13} & 0 & 1.27 & \textbf{98.73} \\
$T_{84}$ & 13 & 7 & 0 & \textbf{100} & 0 & 0 & \textbf{100} & 0 \\
$T_{86}$ & 11 & 6 & 0 & 14.67 & \textbf{85.33} & 0 & 2.33 & \textbf{97.67} \\
$T_{89}$ & 12 & 6 & 0 & \textbf{100} & 0 & 0 & \textbf{100} & 0 \\
$T_{95}$ & 8 & 4 & 0 & 9.93 & \textbf{90.07} & 0 & 1.2 & \textbf{98.8} \\
$T_{96}$ & 15 & 8 & 26.67 & \textbf{73.33} & 0 & 26.67 & \textbf{73.33} & 0 \\
$T_{97}$ & 15 & 8 & 16.67 & \textbf{83.33} & 0 & 16.67 & \textbf{83.33} & 0 \\ \bottomrule
\end{tabular}
}
\end{table*}

\begin{table*}[ht]
\centering
\caption{Comparative analysis of EV charging impact at 7.2 kW and 11.5 kW of power across transformers with 50 kVA capacity in March}
\label{tab: March_Cap50}
\renewcommand{\arraystretch}{1} 
\resizebox{0.8\textwidth}{!}{
\begin{tabular}{lccccccccc}
\toprule
Trans Code & \# Cust & \# EV & \multicolumn{3}{c}{Charging 7.2 kW (\%)} & \multicolumn{3}{c}{Charging 11.5 kW (\%)} \\
\cmidrule(lr){4-6} \cmidrule(lr){7-9}
& & & Infeasibility (\%) & Less \#EV & Desired \#EV & Infeasibility & Less \#EV & Desired \#EV \\
\midrule
$T_{6}$ & 6 & 3 & 0 & 8.27 &\textbf{ 91.73} & 0 & 0.73 & \textbf{99.27} \\
$T_{12}$ & 7 & 4 & 0 & 11.53 &\textbf{ 88.47} & 0 & 1.27 & \textbf{98.73} \\
$T_{13}$ & 7 & 4 & 0 & 8.6 & \textbf{91.4} & 0 & 1.93 & \textbf{98.07} \\
$T_{14}$ & 8 & 4 & 0 & 7.8 & \textbf{92.2} & 0 & 1.27 & \textbf{98.73} \\
$T_{19}$ & 5 & 3 & 0 & 7.33 & \textbf{92.67} & 0 & 0.87 & \textbf{99.13} \\
$T_{30}$ & 6 & 3 & 0 & 6.07 & \textbf{93.93} & 0 & 1.13 & \textbf{98.87} \\
$T_{32}$ & 9 & 5 & 0 & 11.67 & \textbf{88.33} & 0 & 1.93 & \textbf{98.07} \\
$T_{37}$ & 10 & 5 & 0 & 11.73 & \textbf{88.27} & 0 & 1.53 & \textbf{98.47} \\
$T_{40}$ & 8 & 4 & 0 & 12.4 & \textbf{87.6} & 0 & 1.53 & \textbf{98.47} \\
$T_{43}$ & 5 & 3 & 0.07 & 7.26 & \textbf{92.67} & 0 & 1.2 & \textbf{98.8} \\
\bottomrule
\end{tabular}
}
\vspace{-0.5em}
\end{table*}

\begin{table*}[ht]
\centering
\caption{Comparative analysis of EV charging impact at 7.2 kW and 11.5 kW of power across transformers with 75 kVA capacity in March}
\label{tab: March_Cap75}
\renewcommand{\arraystretch}{1} 
\resizebox{0.8\textwidth}{!}{
\begin{tabular}{lccccccccc}
\toprule
Trans Code & \# Cust & \# EV & \multicolumn{3}{c}{Charging 7.2 kW (\%)} & \multicolumn{3}{c}{Charging 11.5 kW (\%)} \\
\cmidrule(lr){4-6} \cmidrule(lr){7-9}
& & & Infeasibility (\%) & Less \#EV & Desired \#EV & Infeasibility & Less \#EV & Desired \#EV \\
\midrule
$T_{54}$ & 9 & 5 & 0 & 11.47 & \textbf{88.53} & 1.53 & 0 & \textbf{98.47} \\
$T_{57}$ & 8 & 4 & 0 & 10.93 & \textbf{89.07} & 1.47 & 0 & \textbf{98.53} \\
$T_{58}$ & 11 & 6 & 0 & 12.47 & \textbf{87.53} & 2 & 0 & \textbf{98} \\
$T_{60}$ & 13 & 7 & 0 & 15.87 & \textbf{84.13} & 2.27 & 0 & \textbf{97.73} \\
$T_{62}$ & 15 & 8 & 0 & 19.73 & \textbf{80.27} & 2.73 & 0 & \textbf{97.27} \\
$T_{66}$ & 9 & 5 & 0 & 12 & \textbf{88} & 2.4 & 0 & \textbf{97.6} \\
$T_{69}$ & 10 & 5 & 0 & 11.07 & \textbf{88.93} & 1.53 & 0 & \textbf{98.47} \\
$T_{77}$ & 10 & 5 & 0 & 12.13 & \textbf{87.87} & 0 & 2.53 & \textbf{97.47} \\
$T_{78}$ & 14 & 7 & 0 & 17.67 & \textbf{82.33} & 0 & 2.53 & \textbf{97.47 }\\
$T_{83}$ & 12 & 6 & 0 & 11.4 & \textbf{88.6} & 0 & 1.93 & \textbf{98.07} \\
$T_{84}$ & 13 & 7 & 0 & 16.73 & \textbf{83.27} & 0 & 2.33 & \textbf{97.67} \\
$T_{86}$ & 11 & 6 & 0 & 15.33 & \textbf{84.67} & 0 & 2.13 & \textbf{97.87} \\
$T_{89}$ & 12 & 6 & 0 & 15.47 & \textbf{84.53} & 0 & 1.8 & \textbf{98.2} \\
$T_{95}$ & 8 & 4 & 0 & 9.8 & \textbf{90.2} & 0 & 1.33 & \textbf{98.6}7 \\
$T_{96}$ & 15 & 8 & 0 & 19.4 & \textbf{80.6} & 0 & 3.2 & \textbf{96.8} \\ 
$T_{97}$ & 15 & 8 & 0 & 18.2 & \textbf{81.8} & 0 & 2.73 & \textbf{97.27 }\\

\bottomrule
\end{tabular}
}
\end{table*}

Consider the row labeled \(T_{6}\) in Table \ref{tab: July_Cap50} for an example of how the data is presented. This transformer serves 6 households implying that the desired number of supported EVs is 3.
Under the 7.2 kW charging scenario, the table lists no instances of infeasibility (0\%), indicating that all households could potentially charge one EV. However, in 6.27\% of scenarios, fewer than the desired three EVs could be charged, while in 93.73\% of the cases, the model successfully generated viable charging schedules that accommodated all 3 EVs. This suggests that for most scenarios under this charging power, the transformer \(T_{6}\) can handle the load of three EVs without any issues.

\subsection{General Observations and Comparative Analysis}
\label{Compar}

In this subsection, the key insights and observations derived from the simulation results (Tables \ref{tab: July_Cap50}-\ref{tab: March_Cap75}) are discussed, focusing on the impact of EV charging across different months and transformers.

\subsubsection{Seasonal Variations in EV Hosting Capacity (HC)}
One notable trend is the greater success rate for accommodating a higher number of EVs in March compared to July. This pattern is particularly evident when comparing transformers \(T_{96}\) and \(T_{97}\) in Tables \ref{tab: July_Cap75} and \ref{tab: March_Cap75}. The improved performance in March can be linked to a lower overall transformer load, as shown by the historical AMI data in Figs. \ref{fig:DailyJuly} and \ref{fig:DailyMarch}. Conversely, the higher loads experienced in July frequently compromise the system's ability to meet the desired EV HC, highlighting the impact of increased seasonal demand.

\subsubsection{Transformer Performance Variability}
Transformer performance varies notably, particularly evident in transformers \(T_{96}\) and \(T_{97}\) from Table \ref{tab: July_Cap75}, which show lower success rates. 
This variability is mainly due to the large numbers of customers (in comparison to their capacity) that these two transformers serve which significantly impacts their potential EV HC. The difference in performance between \(T_{96}\) and \(T_{97}\) which have the same capacities and customer numbers highlight the need for individual asset-level evaluations. 

Additionally, the infeasibility ratios for \(T_{32}\) and \(T_{37}\) in Table \ref{tab: July_Cap50} show significant differences, even though the only variable is that \(T_{37}\) serves one additional household compared to \(T_{32}\). 
\textcolor{black}{A more stark contrast exists between \(T_{62}\) and \(T_{96}\), \(T_{97}\) in Table \ref{tab: July_Cap75}. For instance, \(T_{62}\) and \(T_{96}\) have the same capacity (75 kVA) and number of customers (\#15). However, the coordination model indicates that they have very different capabilities when it comes to hosting EVs: using the proposed coordination strategy, \(T_{62}\) could support the desired number of EVs (namely, \#8) on 76.67\% of the tested scenarios, while \(T_{96}\) could not support \#8 EVs for \textit{any} of the tested scenarios even after coordination.
Such observations highlight the importance of considering unique historical usage data and customer patterns 
to optimize each transformer's performance.
Given these observations, it is recommended that transformers \(T_{37}\), \(T_{96}\), and \(T_{97}\) of this SRP feeder should be prioritized for upgrades to enhance their EV HC.
}

\subsubsection{Impact of Different Charging Power -- 7.2 kW vs. 11.5 kW}
A detailed analysis of different charging levels reveals an interesting pattern: the HC at the higher charging level of 11.5 kW is generally \textit{better} than at 7.2 kW. This is contrary to the initial observation outlined in Section \ref{perform_coordination}, which suggests that lower charging power could better accommodate more EVs due to reduced strain on transformers. For instance, \(T_{62}\) in Table \ref{tab: March_Cap75} exhibits a success rate of 97.27\% for the desired number of EVs at 11.5 kW, which drops significantly to 80.27\% at 7.2 kW. 
\textit{This revised observation
stems from a more realistic modeling of commuting patterns, moving away from the static assumptions such as EV unavailability during typical work hours (8 AM to 4 PM).} 
Note that the 
case-studies considered in Sections \ref{Feeder} personalized commuting behaviors by using the developed joint PMF (see Fig. \ref{fig: DailyCom}). The distributions in that figure (Fig. \ref{fig: DailyCom}) showed varied vehicle availability that do not necessarily coincide with peak electricity demand hours. This results in shorter, more sporadic charging windows and  consequently, the higher charging power of 11.5 kW proves more effective, since it enables faster charging during these limited windows.

However, this comes with a challenge: while higher charging power enhances HC, it simultaneously pushes transformers closer to their operating limits, potentially leading to long-term wear. Therefore, a trade-off lies in balancing the maximization of HC with the minimization of strain on the electric power infrastructure. These insights underscore the complexity of integrating EV charging into residential settings and emphasize the
importance of incorporating dynamic user behavior and realistic commuting patterns into effective grid planning and optimization strategies.


\noindent \textcolor{black}{\textit{Remark:} Note that this study focuses exclusively on residential (home) charging, where the EVs are assumed to be stationary and connected to their home chargers after completing daily trips. 
The impact of traffic-related factors such as congestion, travel speed, topography are \textit{implicitly} covered in the proposed formulation through the worst-case assumption that every EV requires a daily charge up to at least 80\% of its capacity. 
While real-world travel behavior may vary e.g., congestion or longer commutes could lead to higher discharging of the battery,
and on-route charging could reduce home demand, the framework already assumes the most demanding scenario where all EVs charge at home every day. Thus, if transformer limits remain satisfied under the considered extreme condition, they would also hold under more realistic driving and charging patterns.}

\section{Conclusion}
\label{Conclusion}

The study has illuminated critical insights into the EV hosting \textcolor{black}{capacity} of residential power grids under varying operating conditions. 
The effectiveness of the proposed coordination algorithm was evident since it was able to manage substantial EV integration without requiring immediate transformer upgrades, while also pinpointing units that should be given priority (for upgrading) in the future. 
Seasonal dynamics were also found to play a pivotal role, with the month of March demonstrating superior EV load accommodation compared to July. This observation reflects the impact that broader consumption patterns and environmental factors have on EV HC. 
Moreover, in contrast to expectations, higher charging power levels (11.5 kW) notably enhanced HC over lower charging power levels (7.2 kW), aligning closely with the current trend of developing higher-power chargers.
Lastly, the variability in transformer performance across simulations highlighted the intricate interplay of transformer capacity, customer preferences,
and load characteristics, advocating for comprehensive transformer assessments.

In summary, the proposed coordination mechanism offers a systematic approach for utilities to manage EV loads dynamically, ensuring transformer health and system reliability, while also accommodating the future growth of EV usage.
The findings of this case study underscore the necessity of behavioral modeling and targeted infrastructure upgrades to sustainably support increasing EV adoption.
\textcolor{black}{The future scope of this work will involve analyzing the impact of the proposed EV charging coordination scheme on voltage regulation.}

\bibliographystyle{ieeetr}
{\footnotesize
\bibliography{cas_ref.bib}}

\vfill

\end{document}